\documentclass{amsproc}
\newif\ifpdf
 \ifx\pdfoutput\undefined
 \pdffalse
 \else
 \pdfoutput=1
 \pdftrue

 \fi
 \ifpdf
 \usepackage[pdftex]{graphicx}
 \else
\usepackage{graphicx}
 \fi

\newtheorem{theorem}{Theorem}[section]

\theoremstyle{remark}

\numberwithin{equation}{section}

\begin{document}

\title{Gauge Theory in Two Dimensions: Topological, Geometric and Probabilistic Aspects}

\author{Ambar N. Sengupta}
 
\address{Department of Mathematics, Louisiana State University, Baton
Rouge, Louisiana 70803}

\email{sengupta@math.lsu.edu}
\urladdr{http://www.math.lsu.edu/\symbol{126}sengupta}

\thanks{Research supported by US NSF grant   DMS-0601141}

\date{July, 2007.}


\begin{abstract} We present a description of two dimensional Yang-Mills gauge theory
on the plane and on compact surfaces, examining the topological,
geometric and
 probabilistic aspects.

\end{abstract}

\maketitle

\section{Introduction}\label{intro}

  Two dimensional Yang-Mills theory has proved to  be a surprisingly rich model, despite, or possibly
because of, its simplicity and tractability both in classical and
quantum forms. The purpose of this article is to give a largely
self-contained introduction to classical and quantum Yang-Mills
theory on the plane and on compact surfaces, along with its
relationship to Chern-Simons theory, illustrating some of the
directions of current and recent research activity.

\section{Yang-Mills   Gauge Theory }
\label{YMGT}

The physical concept of a gauge field is  modeled mathematically by
the notion of a connection on a principal bundle. In this section we
present a rapid account of the differential geometric notions
describing a gauge field (for a full account, see, for instance,
Bleecker \cite{Bleecker}).

Consider a smooth manifold $M$, to be thought of as spacetime. Let
$G$ be a Lie group, viewed as the group of symmetries of a particle
field.  The latter may be thought of, locally, as a function on $M$
with values in a vector space $E$ on which there is a representation
$\rho$ of $G$; a change of `gauge' (analogous to a change of
coordinates)
 alters $\psi$ by
multiplication by $\rho(g)$, where $g$ is a `local' gauge
transformation, i.e. a function on $M$ with values in $G$. To deal
with such fields in a unified way, it is best to introduce a
principal $G$-bundle over $M$. This is a manifold $P$, with a smooth
surjection
$$\pi:P\to M$$
and a smooth right action of $G$:
$$P\times G\to P:(p,g)\mapsto R_gp= pg,$$
such that $\pi$ is locally trivial, i.e. each point of $M$ has a
neighborhood $U$ for which there is a diffeomeorphism
$$\phi:U\times G\to \pi^{-1}(U)$$
satisfying
$$\pi\phi(u,g)=u,\qquad  \phi(u,gh)=\phi(u,g)h,\qquad\hbox{for all $u\in U$, and $g,h\in G$.}$$
It will be  convenient for later use  to note here that a bundle
over $M$  is   specified concretely by an indexing set $I$ (which
may have a structure, rather than just be an abstract set), an open
covering $\{U_{\alpha}\}_{\alpha\in I}$ of $M$, and for each
$\alpha, \beta\in I$ for which $U_\alpha\cap U_\beta\neq\emptyset$,
a diffeomorphism
$$\phi_{\alpha \beta}:U_{\alpha}\cap U_{\beta}\to G$$
   such that
\begin{equation}\label{phicocyclecondition}
\phi_{\alpha\beta}(x)\phi_{\beta\gamma}(x)=\phi_{\alpha\gamma}(x)
\quad\hbox{for
all $x\in U_{\alpha}\cap U_\beta\cap U_\gamma$.} \end{equation} The
principal bundle $P$ may be recovered or constructed from this data
by taking the set $\cup_{\alpha\in I}\{\alpha\}\times
U_{\alpha}\times G$ and identifying $(\alpha,x,g)$ with $(\beta, y,
h)$ if $x=y\in U_{\alpha}\cap U_{\beta}$ and
$\phi_{\beta\alpha}(x)g=h$. The elements of $P$ are the equivalence
classes $[\alpha,x,g]$, and the map $\pi:P\to M: [\alpha,x,g]\to x$
is  the bundle projection and $[\alpha,x,g]k=[\alpha,x,gk]$
specifies the right $G$-action on $P$. The map $\phi_{\alpha}:
U_{\alpha}\times G\to P:(x,g)\mapsto [\alpha,x,g]$ is a local
trivialization.  This construction is traditional (see, for
instance, Steenrod \cite{Steen}), but lends itself to an interesting
application in the context of Yang-Mills as we shall see later.

 A particle field is then described by a function $\psi:P\to E$, where $E$ is as before, satisfying
 the equivariance property
\begin{equation}\label{sectionpsi}
\psi(pg)=\rho(g^{-1})\psi(p)\end{equation} which is physically
interpreted as the gauge transformation behavior of the field
$\psi$.  In terms of local trivializations, the value of the field
over a point $x\in M$ would be described by  an equivalence class
$[\alpha, x,v]$ with $v=\psi([\alpha,x,e])\in E$, and $(\alpha,x,v)$
declared equivalent to $(\beta,y,w)$ if $y=x\in U_{\alpha}\cap
U_{\beta}$ and $w=\rho\bigl(\phi_{\beta\alpha}(x)\bigr)v$. An
equivalent point of view is to consider the space $E_P$ of all
equivalence classes $[p,v]\in P\times E$, with
$[p,v]=[pg,\rho(g)^{-1}v]$ for all $(g,p,v)\in G\times P\times E$,
which is a vector bundle $E_P\to M: [p,v]\mapsto\pi(p)$, and then
$\psi$ corresponds to the section of this bundle given by $M\to
E_P:x\mapsto [p,\psi(p)]$ for any $p\in\pi^{-1}(x)$. These
traditional considerations will be useful in an unorthodox context
later in defining the Chern-Simons action (\ref{eikCS}).

The interaction of the particle field with a gauge  `force' field is
described through a Lagrangian, which involves derivatives of
$\psi$. This derivative is a `covariant derivative', with a behavior
under gauge transformations controlled through a field $\omega$
which is the gauge field. Mathematically, $\omega$ is a {\em
connection} on $P$, i.e. a smooth
 $1$-form on $P$ with values in the Lie algebra $LG$ of $G$ such that
\begin{equation}\label{Rgomega}
R_g^*\omega={\rm Ad}(g^{-1})\omega\quad\hbox{and}\quad
\omega(pH)=H\end{equation} for all $p\in P$ and $H\in LG$, where
$$pH=\frac{d}{dt}\Big|_{t=0}p\cdot {\rm exp}(tH).$$
The tangent space $T_pP$ splits into the {\em vertical subspace}
$\ker d\pi_p$ and the horizontal subspace $\ker\omega_p$:
\begin{equation}\label{HorVert}
T_pP=V_p\oplus H^{\omega}_p,\quad \mbox{where $V_p=\ker d\pi_p$  and
$H^\omega_p=\ker\omega_p$.}\end{equation} A path in $P$ is said to
be horizontal, or {\em parallel}, with respect to $\omega$, if its
tangent vector is horizontal at every point. Thus, if
$$c:[a,b]\to M$$
is a piecewise smooth path, and $u$ a point on the initial fiber
$\pi^{-1}(c(a))$, then there is a unique piecewise smooth path
${\tilde c}_u:[a,b]\to P$ such that
$$\pi\circ {\tilde c}_u=c,\qquad {\tilde c}_u(a)=u,$$
and ${\tilde c}_u$ is composed of horizontal pieces.  The path
${\tilde c}_u$ is called the {\em horizontal lift}
 of $c$ through $u$, and ${\tilde c}_u(t)$ is the {\em parallel transport} of $u$ along $c$ up to time $t$.

The point ${\tilde c}_u(b)$ lies over the end point $c(b)$. If $c$
is a loop then there is a unique element $h$
 in $G$ for which
$${\tilde c}_u(b)={\tilde c}_u(a)h.$$
This $h$ is the {\em holonomy} of $\omega$ around the loop $c$,
beginning at $u$, and we denote this
$$h_u(c;\omega).$$
   The property (\ref{Rgomega}) implies that
\begin{equation}\label{hugc}
h_{ug}(c;\omega)=g^{-1}h_u(c;\omega)g.\end{equation} In all cases of
interest, the gauge group $G$ is a matrix group, and we have then
the {\em Wilson loop observable}
\begin{equation}\label{def:Wilson}
 {\rm Tr}\bigl(h(c;\omega)\bigr)\end{equation}
where we have dropped the initial point $u$ as it does not affect
the value of the trace of the holonomy.

If $\rho$ is a representation of $G$ on a vector space $E$,  then
there is induced in the obvious way an `action' of the Lie algebra
$LG$ on $E$, and this allows us to multiply, or `wedge', $E$-valued
forms and $LG$-valued forms. If $\eta$ is an $E$-valued $k$-form on
$P$ then the {\em covariant derivative} $D^\omega\eta$ is the
$E$-valued $(k+1)$--form on $P$ given by
\begin{equation}\label{def:Domegaeta}
D^{\omega}\eta=d\eta +\omega\wedge\eta\end{equation}

The holonomy around a small loop is, roughly, the integral of the
{\em curvature} of $\omega$ over the region enclosed
 by the loop. More technically, the curvature $\Omega^\omega$ is the $LG$-valued $2$-form given by
\begin{equation}\label{def:Omegaomega}
\Omega^\omega=D^\omega\omega= d\omega +\frac{1}{2}[\omega\wedge
\omega]
\end{equation}
(see discussion following (\ref{def:CS3form}) below for explanation
of notation). This $LG$-valued $2$-form is $0$ when evaluated on a
pair of vectors at least one of which is vertical, and is
equivariant:
\begin{equation}\label{RgOmega}R_g^*\Omega^\omega={\rm Ad}(g^{-1})\Omega^\omega.\end{equation}
The connection $\omega$ is said to be {\em flat} if its curvature is
$0$, and in this case holonomies around null-homotopic loops are the
identity.

Now consider an Ad-invariant metric $\langle\cdot,\cdot\rangle$ on
$LG$. This, along with a metric on $M$, induces a metric ${\tilde
g}$ on the bundle $P$ in the natural way. Then we can form the
curvature `squared':
$$\langle\Omega^{\omega},\Omega^{\omega}\rangle={\tilde g}
(\Omega^{\omega},\Omega^{\omega})$$
which, by equivariance of the curvature form and the Ad-invariance
of the metric on $LG$, descends to a well-defined function on the
base manifold $M$. The Yang-Mills action functional is
\begin{equation}\label{defSYM}
S_{\rm YM}(\omega)=
\frac{1}{2g^2}\int_M\langle\Omega^{\omega},\Omega^{\omega}\rangle\,
d{\rm
vol}\end{equation} where the integration is with respect to the
volume induced by the metric on $M$. The parameter $g$ is a physical
quantity which we will refer to as the coupling constant. The
Yang-Mills equations are the variational equations for this action.

A {\em gauge transformation} is a diffeomorphism $$\phi:P\to P$$
which preserves each fiber and is $G$-equivariant:
$$\pi\circ\phi=\phi\quad\hbox{and}\quad R_g\circ\phi=\phi\circ R_g \quad\hbox{for all $g\in G$.} $$
The gauge transformation $\phi$ is specified uniquely through the
function $g_{\phi}:P\to G$, given by
\begin{equation}\label{phipg}
\phi(p)=pg_{\phi}(p),\quad\hbox{for all $p\in P$},\end{equation}
which satisfies the equivariance condition
$$g_{\phi}(ph)=h^{-1}g_{\phi}(p)h,\quad\hbox{for all $p\in P$ and
$h\in G$.}$$ Conversely, if  a smooth function $g:P\to G$ satisfies
$g(ph)=h^{-1}g(p)h$ for all $p\in P$ and $h\in G$, then the map
\begin{equation}\label{phigp}
\phi_g:p\mapsto pg(p)\end{equation} is a gauge transformation.
 The group of all gauge
transformations is usually denoted $\mathcal G$; note that the group
law of composition corresponds to pointwise multiplication:
$g_{\phi\circ \tau}=g_{\phi}g_{\tau}$. If $o\in M$ is a basepoint on
$M$, it is often convenient to consider ${\mathcal G}_o$, the
subgroup of $\mathcal G$, which acts as identity on $\pi^{-1}(o)$.
The group $\mathcal G$ acts on the infinite dimensional affine space
$\mathcal A$ of all connections by pullbacks:
\begin{equation}\label{phiomega}
{\mathcal A}\times {\mathcal G}\to{\mathcal A}:(\omega,\phi)\mapsto
\phi^*\omega =\omega^{g_{\phi}}\stackrel{\rm def}{=}{\rm
Ad}\bigl(g_{\phi}^{-1}\bigr)\omega+g_{\phi}^{-1}dg_{\phi}
\end{equation} The Yang-Mills action functional and physical
observables such as the Wilson loop observables are all gauge
invariant.

Gauge groups of interest in physics are products of groups such as
$U(N)$ and $SU(N)$, for $N\in\{1,2,3\}$. In the case of the
electromagnetic field, $G=U(1)$ and the connection form is $\omega
=i\frac{e}{\hbar}A$, where $A$ is the electromagnetic potential,
$\hbar$ is Planck's constant divided by $2\pi$, and $e$ is the
charge of the particle (electron) to which the field is coupled. The
curvature $\Omega^\omega$ descends to an ordinary $2$-form on
spacetime, and corresponds
  to $ \frac{e}{\hbar}$ times the electromagnetic field strength form $F$.

Moving from the classical theory of the gauge field to the quantum
theory leads to the consideration of functional integrals of the
form
$$\int_{{\mathcal A}}f(\omega)e^{ -S_{\rm YM}(\omega)} D\omega,$$
where $f$ is a gauge invariant function such as the product of
traces, in various representations, of holonomies around loops.  The
integral can be viewed as being over the quotient space ${\mathcal
A}/{\mathcal G}$. Here the base manifold $M$ is now a Riemannian
manifold rather than Lorentzian (for the latter, the functional
integrals are Feynman functional integrals, having an $i$ in the
exponent). More specifically, one would like to compute, or at least
gain an understanding of, the averages:
\begin{equation}\label{Wilsonaverage}
W(C_1,...C_k)= \frac{1}{Z_g}\int_{{\mathcal A}/{\mathcal G}}
\prod_{j=1}^k{\rm Tr}\left(h(C_j;\omega)\right) \,e^{- S_{\rm
YM}(\omega)}[D\omega],\end{equation} with $[D\omega]$ denoting the
formal `Lebesgue measure' on ${\mathcal A}$ pushed down to
${\mathcal A}/{\mathcal G}$. Here the traces may be in different
representations of the group $G$. The formal probability measure
$\mu_g$ on ${\mathcal A}/{\mathcal G}$, or on ${\mathcal
A}/{\mathcal G}_o$, given through
\begin{equation}\label{defmug}
d\mu_g([\omega])=\frac{1}{Z_g}e^{-\frac{1}{2g^2}|\!|\Omega^{\omega}|\!|_{L^2}^2}[D\omega],\end{equation}
is usually called the {\em Yang-Mills measure}.

These integrals can be computed exactly when $\dim M=2$, as we will
describe in the following section, and the Yang-Mills measure then
has a rigorous definition.

\section{Wilson loop integrals in two dimensions}\label{S:Wilson}

The Yang-Mills action is, on the face of it, quartic in the
connection form $\omega$. However, when we pass to the quotient
${\mathcal A}/{\mathcal G}$, a simplification results when the base
manifold $M$ is two dimensional. This is most convincingly
demonstrated in the case $M={\mathbb R}^2$. In this case, for any
connection $\omega$ we can choose, for instance, {\em radial gauge},
a section
$$s_{\omega}:{\mathbb R}^2\to P$$
(a smooth map with $\pi\circ s_{\omega}(x)=x$ for all points
$x\in{\mathbb R}^2$) which maps each radial ray from the origin $o$
into an $\omega$-horizontal curve in $P$ emanating from a chosen
initial point $u\in\pi^{-1}(o)$. Then let $F^{\omega}$ be the
$LG$-valued function on ${\mathbb R}^2$ specified by
\begin{equation}\label{defFomega}\omega\mapsto s_{\omega}^*\Omega^{\omega}=F^{\omega}d\sigma,\end{equation}
where $\sigma$ is the area $2$-form on ${\mathbb R}^2$. Then
$$\omega\mapsto F^\omega$$
identifies ${\mathcal A}/{\mathcal G}_o$ with the linear space of
smooth $LG$-valued functions on ${\mathbb R}^2$ and the Yang-Mills
measure becomes the well-defined Gaussian measure on the space of
functions $F$ given by
\begin{equation}\label{defmugR2}
d\mu_g(F)=\frac{1}{Z_g}e^{-\frac{1}{2g^2}|\!|F|\!|_{L^2}^2}\,DF\end{equation}
This measure lives on a completion of the Hilbert space of
$LG$-valued $L^2$ functions on the plane, and the corresponding
connections are therefore quite `rough'. In particular, the
differential equation defining parallel transport needs to be
reinterpreted  as a stochastic differential equation. The holonomy
$h(C;\omega)$ (basepoint fixed at $u$ once and for all) is then a
$G$-valued random variable. The Wilson loop expectation values work
out explicitly using two facts:
\begin{itemize}
\item If $C$ is a piecewise smooth simple closed loop in the plane $C$ then the holonomy $h(C)$ is a $G$-valued random variable with distribution $Q_{g^2S}(x)dx$, where $S$ is the area enclosed by $C$, and $Q_t(x)$ is the solution of the heat equation
$$\frac{\partial Q_t(x)}{\partial t}=\frac{1}{2}\Delta Q_t(x),\qquad \lim_{t\downarrow 0}\int_G f(x)Q_t(x)\,dx=f(e),$$
for all continuous functions $f$ on $G$, with $dx$ being unit mass
Haar measure on $G$, and $\Delta$ is the Laplacian operator on $G$
with respect to the chosen invariant inner product.
\item If $C_1$ and $C_2$ are simple loops enclosing disjoint planar regions then $h(C_1)$ and $h(C_2)$ are independent random variables.
\end{itemize}
In the simplest case, for a simple closed loop $C$ in the plane,
\begin{equation}\label{WilsonsimpleloopR2}
\int f\bigl(h(C)\bigr)\,d\mu_g =\int_G
f(x)Q_{g^2S}(x)\,dx\end{equation} with $S$ denoting the area
enclosed by $C$. In particular, for the group $G=U(N)$,
\begin{equation}\label{WCsimpleplane}
W_N(C)=e^{-Ng^2S/2}\end{equation} where
$$W_N(C)=\int \frac{1}{N}{\rm Tr}\bigl(h(C)\bigr)\,d\mu_{g}.$$

Now consider the case where $M=\Sigma$, a closed oriented surface
with Riemannian structure.    We will follow L{\'e}vy's development
\cite{Le2} of the discrete Yang-Mills measure.  Let $\pi:{\tilde
G}\to G:{\tilde x}\mapsto x $ be the universal covering of $G$. Let
$\mathbb G$ be a triangulation of $\Sigma$, or a graph, with
$\mathbb V$ the set of vertices, $\mathbb E$ the set of (oriented)
edges, and $\mathbb F$ the set of faces. We assume that each face is
diffeomorphic to the unit disk, and the boundary of each face is a
simple loop in the graph. Following L{\'e}vy \cite{Le2}, define a
discrete connection over $\mathbb G$ to be a map ${\tilde
h}:{\mathbb E}\to {\tilde G}$ satisfying
\begin{equation}\label{pietilde}
\pi\bigl({\tilde h}(e^{-1})\bigr)=\pi\bigl({\tilde
h}(e)\bigr)^{-1}\qquad\hbox{for every edge $e\in\mathbb
E$}\end{equation} where $e^{-1}$ denotes the edge $e$ with reversed
orientation. One should interpret ${\tilde h}(e)$ as   the parallel
transport along edge $e$ of a continuum connection lifted to $\tilde
G$ appropriately. Let
$${\mathcal A}_{\mathbb G}$$
be the set of all such connections over $\mathbb G$.  Note that this
is naturally a subset of ${\tilde G}^{\mathbb E}$, and indeed can be
viewed as ${\tilde G}^{{\mathbb E}_+}$, where ${\mathbb E}_+$ is the
set of edges each counted only once with a particular chosen
orientation; in particular, we have a unit mass Haar product measure
on ${\mathcal A}_{\mathbb G}$. Define the {\em discrete Yang-Mills
measure} $\mu^{\rm YM}_g$  for the graph ${\mathbb G}$, by requiring
that for any continuous function $f$ on ${\mathcal A}_{\mathbb G}$,
we have
\begin{equation}\label{discreteYMmeasure}
\int_{ {\mathcal A}_{\mathbb G} }f\,d\mu^{\rm
YM}_g=\frac{1}{Z_g}\int f(h) \prod_{F\in {\mathbb
F}}Q_{g^2|F|}\bigl({\tilde h}(\partial F)\bigr)\,dh,
\end{equation}
where $|F|$ is the area of the face $F$ according to the Riemannian
metric on $\Sigma$, and $Z_g$ a normalizing constant to ensure that
$\mu^{\rm YM}_g({\mathcal A}_{\mathbb G})$ is $1$. This is the
discrete Yang-Mills measure for connections over all principal
${\tilde G}$-bundles over $\Sigma$. However, when $G$ is not simply
connected there are different topological classes of bundles, each
specified through an element  $z\in \ker({\tilde G}\to G)$. For such
$z$, again following L{\'e}vy \cite{Le2},
\begin{equation}\label{ATz}
{\mathcal A}_{\mathbb G}^z=\{{\tilde h}\in {\mathcal A}_{\mathbb
G}\,:\, \prod_{e\in {\mathbb E}_+}{\tilde h}(e){\tilde
h}(e^{-1})=z\}\end{equation} corresponds to the set of connections
on the principal $G$-bundle over $\Sigma$ classified topologically
by $z$. The Yang-Mills measure $\mu^{\rm YM}_{z,g}$ on ${\mathcal
A}_{\mathbb G}^z$ is then simply
\begin{equation}\label{YMmeasurezg}
d\mu^{\rm YM}_{z,g}({\tilde h})= c_z{ 1}_{{\mathcal A}_{\mathbb
G}^z}({\tilde h})d\mu^{\rm YM}_g({\tilde h}),\end{equation} where
$c_z$ is again chosen to normalize the measure to have total mass
$1$. A key feature of the discrete Yang-Mills measure is that it is
unaltered by subdivision of faces (plaquettes), which is why we do
not need to index $\mu^{\rm YM}_g$ by the graph $\mathbb G$; this
invariance was observed by Migdal \cite{Mig1} in the physics
literature. L{\'e}vy \cite{Le1,Le2}    constructed a continuum
measure from these discrete measures and showed that the continuum
measure thus constructed agrees with that constructed in \cite{Se3}.
The continuum construction of the Yang-Mills measure relies on
earlier work by Driver \cite{Dr} and others \cite{GKS}; a separate
approach to the continuum Yang-Mills functional integral in two
dimensions was developed by Fine \cite{Fi1,Fi2} (see also Ashtekar
et al. \cite{ALM}).

The normalizing factor which appears in the loop expectation values
is given, for a simply connected group $G$ and a closed oriented
surface of genus $\gamma$, by
\begin{equation}\label{Zgamma}
 \int_{G^{2\gamma}}Q_{g^2S}\bigl(K_{ \gamma}(x)\bigr)\,dx\end{equation}
where $K_\gamma$ is the product commutator function
\begin{equation}\label{Kgamma}
K_{\gamma}(a_1,b_1,...,a_{\gamma},b_{\gamma})=b_{\gamma}^{-1}a_{\gamma}^{-1}b_{\gamma}a_{\gamma}...b_1^{-1}b_1^{-1}b_1a_1\end{equation}
which plays the role of `total curvature' of a discrete connection
whose holonomies around    $2\gamma$ standard generators of
$\pi_1(\Sigma)$ are given by $a_1,b_1,...,a_\gamma, b_\gamma$.

\section{Yang-Mills on surfaces and Chern-Simons: the symplectic limit}\label{S:YM2}

In this section we will describe   how Yang-Mills theory on surfaces
fits into a hierarchy of topological/geometric field theories in low
dimensions. For a detailed development of Chern-Simons theory from
the point of view of topological field theory we refer to Freed
\cite{Freed} from which we borrow many ideas, and some notation,
here. Most of our discussion below applies to trivial principal
bundles (see \cite{Freed2} for   non-trivial bundles). In Albeverio
et al. \cite{AHS}, the relationship between the Chern-Simons and
Yang-Mills systems was explored using the method of exterior
differential systems of Griffiths \cite{Griff} in the calculus of
variations.

 One of our purposes here is to also verify that the `correct' (from the Chern-Simons point of view) inner-product on the Lie algebra of the gauge group $SU(N)$ to use for two-dimensional Yang-Mills is independent of $N$. This is a small  but
significant fact when considering the large $N$ limit of the
Yang-Mills theory.

\subsection{From four dimensions to three: the Chern-Simons form}

Let $P_W\to W$ be a principal $G$-bundle over a manifold $W$. Then
for any connection $\omega$ on $P_W$ we have the curvature $2$-form
$\Omega^{\omega}$ which gives rise to an $LG\otimes LG$-valued
$4$-form by wedging
$$\Omega^{\omega}\wedge \Omega^{\omega}$$
Now consider a metric $\langle\cdot,\cdot\rangle$ on $LG$ which is
Ad-invariant. This produces a  $4$-form
$$\langle \Omega^{\omega}\wedge \Omega^{\omega}\rangle$$
which,   by Ad-invariance, descends to a $4$-form on $W$ which we
denote again by $\langle \Omega^{\omega}\wedge
\Omega^{\omega}\rangle$. The latter, a Chern-Weil form, is a closed
$4$-form and specifies a cohomology class in $H^4(W)$ determined by
the bundle $P_W\to W$ (independent of the choice of $\omega$).

The {\em Chern-Simons} $3$-form $cs(\omega)$ on the bundle space
$P_W$ is given by
\begin{equation}\label{def:CS3form}
cs(\omega)=\langle \omega\wedge d\omega +\frac{1}{3}\omega\wedge
[\omega\wedge \omega]\rangle =\langle \omega\wedge\Omega^{\omega
}-\frac{1}{6}\omega\wedge [\omega\wedge \omega]\rangle\end{equation}
Here wedge products of $LG$-valued forms, and expressions such as
$[\omega\wedge\omega ]$, may be computed by expressing the forms in
terms of a basis of $LG$ and ordinary differential forms. For
example, writing $\omega$ as $\sum_a\omega^aE_a$, where $\{E_a\}$ is
a basis of $LG$,  the $2$-form
 $[\omega\wedge\omega ]$,  whose value on a pair of vectors
$(X,Y)$ is $2[\omega(X), \omega(Y)]$, is $\sum_{a,b}\omega^a\wedge
\omega^b [E_a,E_b]$. If $LG$ is realized as a Lie algebra of
matrices, then $[\omega\wedge\omega]$ works out to be
$2\omega\wedge\omega$, this being computed using matrix
multiplication.

The fundamental property    \cite{CS}  of the Chern-Simons form is
that its exterior differential is the closed $4$-form
$\langle\Omega^{\omega}\wedge \Omega^{\omega}\rangle$ on the bundle
space:
\begin{equation}\label{dCSomega}
dcs(\omega)= \langle \Omega^{\omega}\wedge
\Omega^{\omega}\rangle\end{equation} Unlike the Chern-Weil form,
$cs(\omega)$ does not descend naturally to a form on $W$, i.e. if
$s:W\to P_W$ is a
 section then $s^*cs(\omega)$ depends on $s$. If $g:P\to G$ specifies a gauge transformation $p\mapsto pg(p)$
 then a lengthy but straightforward computation shows that
 \begin{equation}\label{CSomegag}
cs(\omega^g)-cs(\omega)= d\langle \omega\wedge (dg)g^{-1}\rangle
-\frac{1}{6}\langle g^{-1}dg\wedge [g^{-1}dg\wedge
g^{-1}dg]\rangle\end{equation}

If we split  a closed oriented $4$-manifold $W$   into two
$4$-manifolds $W_1$ and $W_2$, glued along a compact oriented
$3$-manifold $Y$, and if  $P_W$ admits sections $s_1$ over $W_1$ and
$s$ over $W_2$, then
\begin{equation}\label{OmegawCSo}
\int_W \langle \Omega^{\omega }\wedge \Omega^{\omega }\rangle=\int_Y
\bigl(s_1^*cs(\omega)- s^*cs(\omega)\bigr)\end{equation} Now the
sections $s_1$ and $s$ are related by a gauge transformation $g$
specified through a smooth map
\begin{equation}\label{tildgY}
{\tilde g}:Y\to G\end{equation} in the sense that (notation as in
(\ref{phipg}) and (\ref{phigp}))
\begin{equation}\label{intYgsig}s_1(y)=s(y){\tilde
g}(y)=\phi_g(s(y)),\qquad\hbox{for all $y\in Y$.}\end{equation}

 Then, after using Stokes' theorem, the term on the right in (\ref{OmegawCSo}) works out
to
\begin{equation}\label{intYgsigma}-\frac{1}{6}\int_Y {\tilde
g}^*\sigma \end{equation} where $\sigma$ is the $3$-form on $G$
given by
\begin{equation}
\sigma=\langle h^{-1}dh\wedge [h^{-1}dh\wedge h^{-1}dh]
\rangle,\label{defsigma}\end{equation} with $h:G\to G$ being the
identity map. By choosing the metric on $LG$ appropriately, we can
ensure that this quantity is always an integer times (a convenient
normalizing factor) $8{\pi^2}$.  For example, if $G=SU(2)$, and the
inner-product on $LG$ given by
\begin{equation}\label{lHKr}\langle H,K\rangle =- {\rm
Tr}(HK),\end{equation} computation of the volume of $SU(2)$ shows
that
\begin{equation}
\int_{SU(2)}\sigma=-48\pi^2\end{equation} (The sign on the right
just fixes an orientation for $SU(2)$.)  This computation can be
worked out conveniently through the 2-to-1 parametrization of
$SU(2)$ given by $h = k_\phi a_\theta k_\psi$, with $(\phi,\theta,
\psi)\in (0,2\pi)\times (0,\pi)\times (0,2\pi)$, where
$$k_t =    \left(\begin{matrix} e^{i{t}}&0\\ 0& e^{-i{t}}\end{matrix}\right)
$$
and
$$a_\theta
 =   \left(\begin{matrix}  \cos{\frac{\theta}{2}}& i\,
\sin{\frac{\theta}{2}}\\
i\,   \sin{\frac{\theta}{2}}&
\cos{\frac{\theta}{2}}\end{matrix}\right)
$$
Putting all this together we see that
\begin{equation}\label{CSomegag8pi}
\int_W\left[\frac{1}{8\pi^2}s^*cs(\omega^g)-\frac{1}{8\pi^2}s^*cs(\omega)\right]=-\int_Y
\frac{1}{48\pi^2}{\tilde g}^*\sigma\in {\mathbb Z}\end{equation}

More generally, we assume that the metric
$\langle\cdot,\cdot\rangle$ is  such that
$$\frac{1}{8\pi^2}\langle \Omega^{\omega}\wedge \Omega^{\omega}\rangle$$ is an integer
cohomology class for every closed oriented four-manifold $W$ (this
condition can be restated
 more completely in terms of the classifying space $B{\tilde G}$; see Witten \cite{Wi1}). For instance,
for $G=SU(N)$, the properly scaled metric is (according to Witten
\cite{Wi1}):
\begin{equation}\label{defmetricSU2}\langle H,K\rangle =-{\rm Tr}(HK)\end{equation}
Let
\begin{equation}\label{CSaction}
CS(s,\omega)=\frac{1}{8\pi^2}\int_Y s^*cs(\omega)\end{equation}
where $s:Y\to P$ is a smooth global section (assumed to exist); the
discussions above show that when $s$ is altered, $CS(\omega)$ is
changed by an integer. Thus, for any integer $k\in\mathbb Z$, the
quantity
\begin{equation}
e^{2{\pi}ki CS(s,\omega)}\in U(1)\end{equation} is independent of
the   section $s$, and thus gauge invariant.
\subsection{From three dimensions to two: the $U(1)$ bundle over the space of connections on a surface}

Now consider a compact oriented $3$-manifold $Y$ with boundary $X$,
a closed oriented $2$-manifold. We follow Freed's approach
\cite{Freed}. We assume that $G$ is  connected, compact, and simply
connected; a consequence is that a  principal $G$-bundle over any
manifold of dimension $\leq 3$ is necessarily trivial. For any
smooth sections $s_1, s:Y\to P$, with $s_1=s{\tilde g}$, we have on
using (\ref{CSomegag}) and notation explained therein,
 \begin{equation}
\int_Y s^*cs(\omega^g)-\int_Ys^*cs(\omega)=\int_X\langle
s^*\omega\wedge (d{\tilde g}){\tilde g}^{-1}\rangle
-\int_Y\frac{1}{6}{\tilde g}^*\sigma
\end{equation}
Let
\begin{equation}\label{CSsomega}
CS(s,\omega)=\frac{1}{8\pi^2}\int_Ys^*cs(\omega)\end{equation} Then,
for any integer $k$,
\begin{equation}\label{CScocycle}
e^{2{\pi}ki CS(sg,\omega)}=e^{2{\pi}ki
CS(s,\omega)}\phi_{sg,s}(\omega)
\end{equation}
where
\begin{equation}\label{sgcocycle}
\phi_{s{\tilde
g},s}(\omega)=e^{2{\pi}ki\left[\frac{1}{8\pi^2}\int_X\langle
s^*\omega\wedge (d{\tilde g}){\tilde g}^{-1}\rangle -
\int_Y\frac{1}{48\pi^2}{\tilde g}^*\sigma\right]}
\end{equation}
The second term in the exponent on the right  is determined, due to
integrality of $\sigma$, by  ${\tilde g}|X$, and is independent of
the extension of  ${\tilde g}$ to $Y$. Thus
\begin{equation*}\mbox
 {\em $\phi_{s{\tilde g},s}(\omega)$ is determined by  $s$, $\omega$, and ${\tilde g}$ on the two-manifold $X$.}
 \end{equation*} These data specify a principal
$U(1)$ bundle over the space ${\mathcal A}_X$ of connections on the
bundle $P_X\to X$ (restriction of $P$ over $X$), as follows. Let $I$
be the set of all smooth sections $s:X\to P_X$. Taking this as
indexing set,   if $s_1,s\in I$ then, denoting by ${\tilde g}:X\to
G$ the function for which $s_1=s{\tilde g}$, we define
$\phi_{s_1,s}$ as above. Thus, $\phi_{s_1,s}(\omega)$ is given by
\begin{equation}\label{defphis1s}
\phi_{s_1,s}(\omega)=e^{2{\pi}ki
CS(s_1,\omega)}e^{-2{\pi}kiCS(s,\omega)},\end{equation} where, on
the right,  the Chern-Simons actions are computed for   extensions
of $s^*\omega$ and $s^*\omega^g$   over a $3$-manifold $Y$ whose
boundary is $X$. If a different $3$-manifold $Y'$ is chosen then
the value of $\phi_{s_1,s}(\omega)$ remains the same, because it
gets multiplied by
$$e^{2{\pi}ki CS_{Y'\cup -Y}(s_1,\omega)}e^{-2{\pi}kiCS_{Y'\cup -Y}(s,\omega)},$$
with obvious notation, and we have seen that this is $1$. The
expression (\ref{defphis1s}) makes it clear that
$\{\phi_{s_1,s}\}_{s,s_1\in I}$ satisfies the cocycle condition
(\ref{phicocyclecondition}) and thus {\em specifies a principal
$U(1)$-bundle over the space ${\mathcal A}_X$ of connections on
$P_X\to X$}.

Note that the integrality condition on $k$ (which goes back to the
integrality property of the inner-product on $LG$) is what leads to
the $U(1)$ bundle.

The principal $U(1)$-bundle constructed along with the natural
representation of $U(1)$ on $\mathbb C$, yields a line bundle
$\mathbb L$ over ${\mathcal A}_X$, as described more generally in
the context of (\ref{sectionpsi}). If $Y$ is a $3$-manifold with
boundary $X$ then for any connection $\omega$ on the bundle over
$Y$, we have a well-defined element
\begin{equation}\label{eikCS}
e^{2{\pi}kiCS(\omega)}\stackrel{\rm def}{=}[s,
e^{2{\pi}kiCS(s,\omega)}]\end{equation} in the $U(1)$-bundle over
${\mathcal A}_X$ in the fiber over $\omega|X$. In this way
(following Freed \cite{Freed}), the {\em exponentiated Chern-Simons
action over an oriented $3$-manifold $Y$ with boundary $X$ appears
as a section of the line bundle $\mathbb L$ over} ${\mathcal A}_X$.

\subsection{Connection on the $U(1)$ bundle over the space of connections}

The method of geometric quantization also requires a connection on
the $U(1)$-bundle (over phase space). The connection is here
generated again using the Chern-Simons action. Let
$$[0,1]\to {\mathcal A}_X: t\mapsto \omega_t $$
be a path of connections, such that $(t,p)\mapsto \omega_t(p)$ is
smooth. Then this specifies a connection $\omega$ on the bundle
$$[0,1]\times P\to [0,1]\times X $$
in the obvious way (parallel transport in the $t$ direction is
trivial). We define   parallel transport along the path
$t\mapsto\omega_t$ over ${\mathcal A}_X$  geometrically as follows:
consider any $3$-manifold $Y$ with boundary $X$, and a principal
$G$-bundle $P_Y\to Y$ with connection $\omega_{0,Y}$ which restricts
to the given bundle over $X$ and $\omega_0$, and similarly  consider
$\omega_{1,Y}$; then parallel-transporting
$e^{2{\pi}kiCS(\omega_{0,Y})}$ along the path will yield
$$e^{2{\pi}kiCS(\omega_{1,Y})}e^{2{\pi}ki CS({\tilde\omega})}$$
where ${\tilde\omega}$ is the connection over $(Y)\cup (X\times
[0,1])\cup (-Y)$, glued along $X$, obtained by combining $\omega$,
$\omega_{0,Y}$ and $\omega_{1,Y}$. In terms of a trivialization of
the bundle specified through a section $s$ of $P$ over $X$, parallel
transport is given by multiplication by
\begin{equation}\label{paralleltransp}
e^{2{\pi}ki CS({\tilde s},{\tilde\omega})}\end{equation} where
$\tilde s$ is the induced trivialization of $[0,1]\times P\to
[0,1]\times X$. Observe that (indicating by the subscript $X$ the
differential over $X$)
$$d{\tilde s}^*{\tilde \omega}=d_Xs^*\omega_t +dt\wedge
\frac{\partial s^*\omega_t}{\partial t}.  $$
A simple computation then shows
\begin{equation}\label{CSaspt}
  CS({\tilde s},{\tilde\omega})=-
  \frac{1}{8\pi^2}\int_{[0,1]}
   \left(\int_X\left\langle \omega_t\wedge
   \frac{\partial\omega_t}{\partial t}\right\rangle\right)\wedge dt.
\end{equation}
Viewing the Lie algebra of $U(1)$ as $i{\mathbb R}$, the parallel
transport for a $U(1)$ connection along a path is $e^{-P}$, where
$P$ is the integral of the connection form along the path, we see
that the connection form $\theta$ on the $U(1)$ bundle  over
${\mathcal A}_X$ is given explicitly by
\begin{equation}\label{omegaAX}
\theta|_\omega (A) = 2{\pi}i \frac{k}{8\pi^2} \int_X\left\langle
\omega\wedge A\right\rangle,
\end{equation}
for any connection $\omega\in {\mathcal A}_X$ and any vector $A$
tangent to ${\mathcal A}_X$ at $\omega$ (such an $A$ is simply an
$LG$-valued $1$-form on $P$ which vanishes on vertical vectors and
satisfies $R_g^*A={\rm Ad}(g^{-1})A$ for every $g\in G$). The
curvature of this is given by the $i{\mathbb R}$-valued $2$-form
$\Theta=d\theta$ specified explicitly on ${\mathcal A}_X$ by
$$\Theta(A,B)=A\bigl(\Theta(B)\bigr)- B\bigl(\Theta(A)\bigr)$$
(where $A$ and $B$ are treated as `constant' vector fields on the
affine space ${\mathcal A}_X$). This yields
\begin{equation}\label{Theta}
\Theta(A,B)= 2{\pi}i \frac{k}{4\pi^2} \int_X\left\langle A\wedge
B\right\rangle
\end{equation}
for all $A,B\in T_{\omega}{\mathcal A}_X$. In keeping with the
Bohr-Sommerfeld quantization conditions, we should consider the the
symplectic form
\begin{equation}\label{2piinvTheta}
\frac{1}{2{\pi}i}\Theta=\frac{k}{4\pi^2} \int_X\left\langle A\wedge
B\right\rangle
\end{equation}
This is precisely, with correct scaling factors, the symplectic
structure used by Witten [equation (2.29) in \cite{Wi1}] with $k=1$.

In the context of geometric quantization it is more common to
consider the Hermitian line bundle associated to the principal
$G$-bundle over ${\mathcal A}_X$ constructed here, and view the
connection as a connection on this line bundle. From this point of
view one might as well simply consider the case $k=1$, since the
case of general $k\in\mathbb Z$ arises from different
representations of $U(1)$, i.e. are tensor powers of the $k=1$ line
bundle (and its conjugate).

\subsection{From Chern-Simons to Yang-Mills on a surface}

 The original gauge invariance of $e^{2{\pi}kiCS(\omega)}$
  transfers to an easily-checked gauge invariance of the
  symplectic structure $\Theta$ on the space of connections.
   Thus, we have
the group ${\mathcal G}$ of all gauge transformations acting
symplectically on the affine space ${\mathcal A}_X$. As is well
known,   this action has a moment map:
\begin{equation}\label{moment}
J:{\mathcal A}_X\to (L{\mathcal G})^*:\omega\mapsto \frac{k}{4\pi^2}
\Omega^{\omega}
\end{equation}
where we have identified the dual of the infinite dimensional Lie
algebra $L{\mathcal G}$ with the space of $LG$-valued Ad-equivariant
functions on the bundle space $P$. This fact is readily checked
using Stokes' theorem:
$$\langle J'(\omega)A, H\rangle =\frac{k}{4\pi^2}
 \int_X \langle (dA+[\omega\wedge A]), H\rangle=\frac{1}{2{\pi}i}\Theta(A, dH+[\omega, H])$$
The {\em Yang-Mills} action now can be seen as the norm-squared of
the moment map:
\begin{equation}\label{SYM}
S_{\rm YM}(\omega)= \frac{1}{2g^2}
|\!|\frac{4\pi^2}{k}J|\!|^2\end{equation} where $|\!|J|\!|^2$ is
computed as an $L^2$-norm squared.

We have been discussing Chern-Simons theory in terms of its action,
i.e. the integral of the Lagrangian. The Hamiltonian picture works
with the {\em phase space}, i.e. the space of extrema of the
Chern-Simons action. A fairly straightforward computation shows that
the extrema are flat connections. If we consider the $3$-manifold
$$Y=[0,T]\times\Sigma,$$
where $\Sigma$ is a closed oriented surface, then the phase space,
after quotienting out the gauge symmetries, may be identified as the
moduli space of flat connections over $\Sigma$, which in turn is
$J^{-1}(0)/{\mathcal G}$. It is a stratified space, with  maximal
stratum ${\mathcal M}^0$ which is a symplectic manifold with
symplectic structure  induced by $\frac{1}{2{\pi}i}\Theta$. We
denote this symplectic structure by $\overline\Omega$ when $k$ is
set to $1$, i.e. it is induced by the symplectic structure on
${\mathcal A}_X$ given by
\begin{equation}\label{2piinvThetak1}
 \frac{1}{4\pi^2} \int_X\left\langle A\wedge B\right\rangle
\end{equation}

\subsection{The symplectic limit}

The formal Chern-Simons path integral $$\int_{{\mathcal A}_Y}
e^{2{\pi}kiCS(\omega)}D\omega$$ is naturally of interest in the
quantization of Chern-Simons theory (for progress on a rigorous
meaning for Chern-Simons functional integrals see Hahn \cite{Hahn1,
Hahn2}). The path integral may be analyzed in the $k\to\infty$ limit
by means of its behavior at the extremal of $CS$, i.e. on the moduli
space   of flat connections. This is also what results when we
examine the limit of the Yang-Mills measure
$$\frac{1}{N_g}e^{-\frac{1}{2g^2}|\!|\Omega^{\omega}|\!|^2}D\omega,$$
(with $N_g$ a formal normalizing factor) for connections over the
surface $X$, in the limit $g\to 0$.

Formally, it is clear that the limiting measure, if it is
meaningful, should live on those connections where $\Omega^{\omega}$
is $0$, i.e. the flat connections. Quotienting by gauge
transformations yields the moduli space ${\mathcal M}^0$ of flat
connections. For a compact oriented surface $\Sigma$ of genus
$\gamma\geq 1$, the fundamental group $\pi_1(\Sigma,o)$, where $o$
is any chosen basepoint, is generated by the homotopy classes of
loops $A_1,B_1,...,A_\gamma,B_\gamma$ subject to the constraint that
the word $B_{\gamma}^{-1}A_{\gamma}^{-1}B_{\gamma}A_{\gamma}\ldots
B_{1}^{-1}A_{1}^{-1}B_{1}A_1$ is the identity in homotopy.
Considering a (compact, connected,) simply connected gauge group $G$
(so that a principal $G$-bundle over $\Sigma$ is necessarily
trivial), each flat connection is specified, up to gauge
transformations, by the holonomies around the loops $A_i$, $B_i$. In
this way, ${\mathcal M}^0$ is then identified with the subset of
$G^{2{\gamma}}$, modulo conjugation by $G$, consisting of all
$(a_1,b_1,...,a_{\gamma},b_{\gamma})$ satisfying
$$b_{\gamma}^{-1}a_{\gamma}^{-1}b_{\gamma}a_{\gamma}
\ldots b_1^{-1}a_1^{-1}b_1a_1=e.$$

 Recalling our description of the Yang-Mills measure in terms
 of the heat kernel $Q_t$ on $G$, we have the following result\cite{Se9}:
\begin{theorem} Consider a closed, oriented Riemannian
 two-manifold of genus $\gamma\geq 2$, and assume that $G$ is a compact, connected, simply-connected Lie group, with Lie algebra equipped with an Ad-invariant metric. Let $f$ be
a $G$-conjugation invariant continuous  function on $G^{2\gamma}$,
and $\tilde f$ the induced function on subsets of $G^{2\gamma}/G$.
Then
\begin{equation}\lim_{t\downarrow 0}\int_{G^{2{\gamma}}}f(x)Q_{t}\bigl(K_{\gamma}(x)\bigr)\,dx= \frac{(2\pi)^{n}}{|Z(G)|{\rm vol}(G)]^{ 2{\gamma}-2}}\int_{{\mathcal M}^0}{\tilde f}\,d{\rm vol}_{\overline\Omega},\end{equation}
where $|Z(G)|$ is the number of elements in the center $Z(G)$ of
$G$, and  ${\rm vol}_{\overline\Omega}$ is the symplectic volume
form $\frac{1}{n!} {\overline\Omega}^{n}$ on the space ${\mathcal
M}^0$ whose dimension is $2n=(2\gamma-2)\dim G$.
\end{theorem}
With $f=1$ this yields Witten's volume formula (see equation (4.72)
in \cite{Wi2})
\begin{equation}
{\rm vol}_{\overline\Omega}\bigl({\mathcal
M}^0\bigr)=\frac{|Z(G)|{\rm vol}(G)]^{
2{\gamma}-2}}{(2\pi)^{n}}\sum_\alpha\frac{1}{(\dim\alpha)^{2\gamma-2}}
\end{equation}
where the sum is over all non-isomorphic irreducible representations
$\alpha$ of $G$. Specialized to $G=SU(2)$, this gives the symplectic
volume of the moduli space of flat $SU(2)$ connections over a closed
genus $\gamma$ surface to be the rational number
$\frac{2^{\gamma-1}}{(2\gamma-2)!}(-1)^{\gamma}B_{2\gamma-2}$, where
$B_k$ is the $k$-th Bernoulli number. (Note that keeping track of
all the factors of $2\pi$ pays off in reaching this rational
number!)

 \section{Concluding Remarks}

 We have given an overview of the geometric and topological aspects of two-dimensional Yang-Mills theory and described how they relate to the Yang-Mills probability measure.

Many physical systems involving a parameter $N$ have asymptotic
limiting forms as $N\to\infty$, even though such a limit may not
have a clear physical meaning. For the case of Yang-Mills gauge
theory with gauge group $U(N)$, the limit as $N\to\infty$ (holding
$g^2N$ fixed, where $g$ is the coupling constant) has been of
particular interest since the path breaking work of `t Hooft
\cite{tHft}. We refer to the recent review \cite{Se13} for more
details on the large $N$ limit of Yang-Mills in two dimensions.  A
key observation is that letting $N\to\infty$, while holding ${\tilde
g}^2=g^2N$ fixed, yields meaningful finite limits of all Wilson loop
expectation values. There is also good reason to believe (see Singer
\cite{Si}) that a meaningful $N=\infty$ theory also exists, possibly
with relevance to Yang-Mills gauge theory in higher dimensions as
well.  Free probability theory (see, for instance, Voiculescu et al.
\cite{VDN} and Biane \cite{Bi}) is likely to play a significant role
here.

The partition function for $U(N)$ gauge theory on a genus $\gamma$
surface is the normalizing constant we have come across:
$$Z_{\gamma}=\sum_{\alpha}(\dim\alpha)^{2-2\gamma}
 e^{-{\tilde g}^2Sc_2(\alpha)/(2N)}$$
where the sum is over all distinct irreducible representations
$\alpha$ of $U(N)$, which may be viewed as a sum over the
corresponding Young tableaux (which parametrize the irreducible
representations), and $c_2(\alpha)$ is the quadratic Casimir for
$\alpha$. This sum may be viewed naturally as a statistical
mechanical partition function for a system whose states are given by
the Young tableaux. This point of view leads to the study of
Schur-Weyl duality for $U(N)$ gauge theory (see, for example,
\cite{AP}) and to the study of phase transitions in the parameter
${\tilde g}^2S$ as $N\uparrow\infty$, viewed as a thermodynamic
limit.

The references below present a sample of relevant works, and does
not aspire to be a comprehensive bibliography.

{\bf Acknowledgment}.   Research support from the U.S. National
Science Foundation   (Grant  DMS-0601141) is gratefully
acknowledged.

\end{document}